\documentclass[submission,copyright,creativecommons]{eptcs}
\providecommand{\event}{Post Proceedings of ADG'23} 
\usepackage[T1]{fontenc}
\usepackage{graphicx}
\usepackage{doi}

\usepackage{xurl}
\usepackage{hyperref}
\begin{document}
\title{\thanks{Second and third authors supported by a grant PID2020-113192GB-I00 (Mathematical Visualization: Foundations, Algorithms and Applications) from the Spanish MICINN.
}}
\def\titlerunning{GeoGebra Discovery and an Austrian Mathematics Olympiad Problem}
%
%
\def\authorrunning{B. Ari\~no-Morera et al.}
%
\title{Solving with GeoGebra Discovery an Austrian Mathematics Olympiad Problem: Lessons Learned}
\author{Bel\'en Ari\~no-Morera
\institute{Departamento de Econom{\'\i}a Financiera y Contabilidad, Universidad Rey Juan Carlos, Madrid, Spain}
\email{belen.arino@urjc.es}
\and
Zolt\'an Kov\'acs
\institute{The Private University College of Education of the Diocese of Linz, Austria}\thanks{The
work was partially supported by the grant 
PID2020-113192GB-I00 from the Spanish MICINN.}
\email{zoltan.kovacs@ph-linz.at}
\and
Tom\'as Recio
\institute{Escuela Polit\'ecnica Superior, Universidad Antonio de Nebrija, Madrid, Spain}
\email{trecio@nebrija.es}
\and
Piedad Tolmos
\institute{Departamento de Econom{\'\i}a Financiera y Contabilidad, Universidad Rey Juan Carlos, Madrid, Spain}
\email{piedad.tolmos@urjc.es}}
\maketitle              
\begin{abstract}
We address, through the automated reasoning tools in GeoGebra Discovery, a problem from a regional phase of the Austrian Mathematics Olympiad 2023. Trying to solve this problem gives rise to four different kind of feedback: the almost instantaneous, automated solution of the proposed problem; the measure of its complexity, according to some recent proposals;  the automated discovery of a generalization of the given assertion, showing that the same statement is true over more general polygons than those mentioned in the problem; and the difficulties associated to the analysis of  the surprising and involved high number of degenerate cases that appear when using the \texttt{LocusEquation} command in this problem.
In our communication we will describe and reflect on these diverse issues, enhancing its exemplar role for showing some of  the advantages, problems, and current fields of development of GeoGebra Discovery.

\end{abstract}

\section{Introduction}
In the past years we have been developing and including, both in the standard version of GeoGebra\footnote{\url{www.geogebra.org}} as well as in the fork version GeoGebra Discovery\footnote{\url{https://kovzol.github.io/geogebra-discovery/}}, different automated reasoning tools (ART) \cite{KRV22}. Mathematics Olympiads problems provide an interesting benchmark for testing the performance of such instruments. Interesting,  from
multiple perspectives:  by itself, as a source of challenging mathematical questions that our ART should be able to deal with; as a test to measure and compare the performance of humans and machines on the same problems \cite{ART},\cite{AM}; as a test on the suitability of the recently proposed intrinsic measure of complexity  of a geometric statement \cite{KRV23}. 
  
\begin{figure}
 \begin{center}
\includegraphics[scale=0.18]{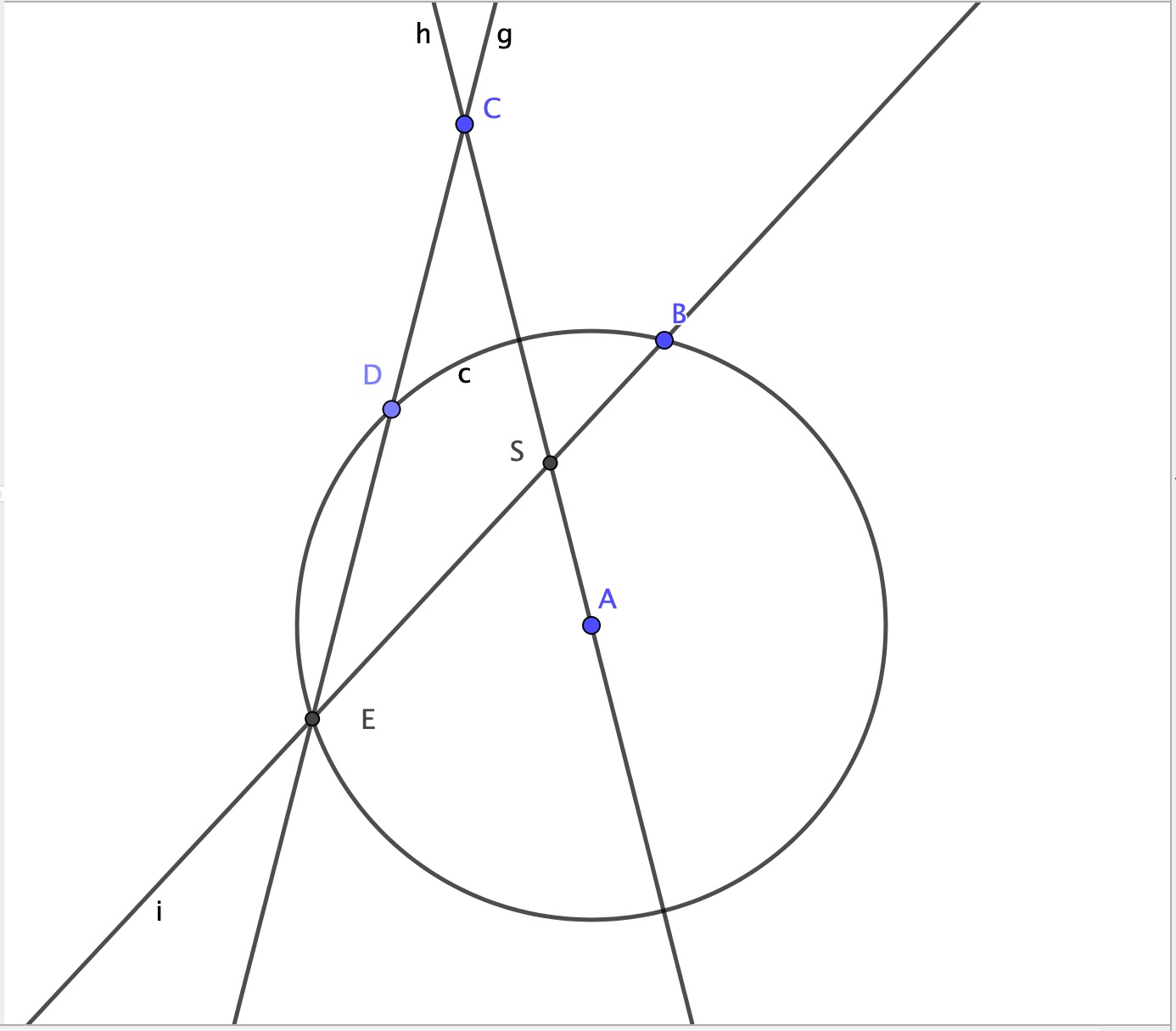}
\caption{Problem 2.  \"Osterreichische Mathematik-Olympiade 
Regionalwettbewerb f\"ur Fortgeschrittene 30. M\"arz 2023.} 
\label{fig1}
\end{center}
\end{figure}

In this communication we will focus on how our ART behave concerning some of these issues, when addressing an interesting problem recently proposed in a regional phase of the Austrian Mathematics Olympiad 2023. Namely,  Problem 2
at the   54.~Austrian Mathematics Olympiad 2023,\footnote{\"Osterreichische Mathematik-Olympiade 
Regionalwettbewerb f\"ur Fortgeschrittene, 30.~M\"arz 2023. \url{https://oemo.at/OeMO/Termine/2023/}}
stated as follows (see Figure \ref{fig1}):
 
 \begin{quotation}
 Sei $ABCD$ eine Raute mit  $\angle BAD < 90^\circ$. Der Kreis durch $D$ mit Mittelpunkt $A$ schneide die Gerade $CD$ ein zweites MaI im Punkt $E$. Der Schnittpunkt der Geraden $BE$ und $AC$ sei $S$.
 
Man beweise, dass die Punkte $A, S, D$ und $E$ auf einem Kreis liegen.
 \end{quotation}

that is

 \begin{quotation}
Let $ABCD$ be a rhombus with $\angle BAD < 90^\circ$. The circle through $D$ with center $A$ intersects straight line $CD$ a second time at point $E$. The intersection of the lines $BE$ and $AC$ is $S$.
 
Prove that the points $A, S, D$ and $E$ lie on a circle.

 \end{quotation}

In the next section we will show how GeoGebra Discovery is able to confirm (by proving internally, in a mathematically rigorous way) the truth of the proposed problem.

\section{Solving the Problem}

\begin{figure}
 \begin{center}
\includegraphics[width=\linewidth]{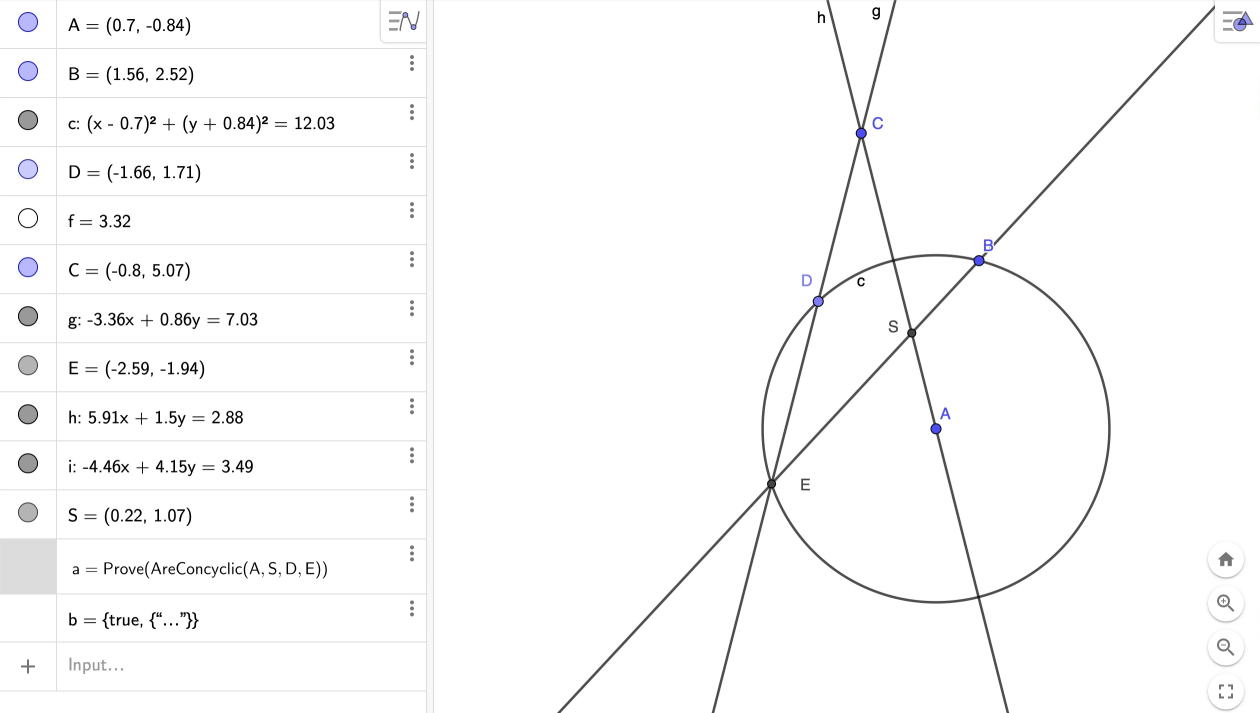}
\caption{Confirming the truth of the statement in Problem 2.} 
\label{fig2}
\end{center}
\end{figure}

In what follows, to exemplify how GeoGebra Discovery handles this problem,  we will use  in different moments,  the web version GeoGebra Discovery 6 or the app GeoGebra Discovery 5. To start with, Figure \ref{fig2} shows how GeoGebra Discovery version 6.0.641.0-2023Apr22  confirms the truth of the given statement.

First, we have chosen some free points $A,B$, then the circle $c$ centered at $A$ through $B$,
then another point $D$ on this circle, such that $\angle BAD < 90^\circ$. Next, we have built the (hidden) segment $f=BD$, and point $C$ as the symmetrical of $A$ with respect to $f$. Thus, $ABCD$ is a rhombus.  Finally, points $E, S$ are displayed, following the hypotheses, as the intersection of line $CD$ and $c$ (ditto, as the intersection of line $BE$ and $AC$).

\begin{figure}
 \begin{center}
\includegraphics[width=\linewidth]{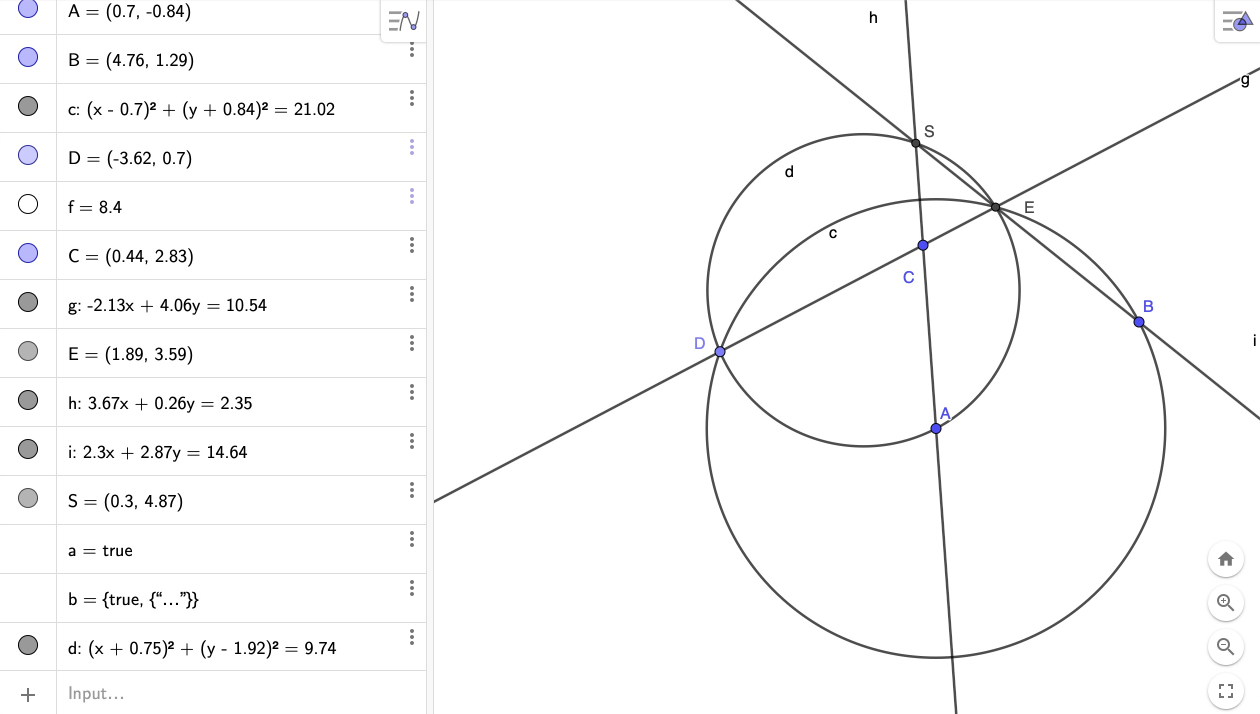}
\caption{Confirming the truth of the statement in Problem 2 and showing the circle through $A,S,D,E$.} 
\label{fig3}
\end{center}
\end{figure}

 Then we have introduced the commands \begin{center}\texttt{Prove(AreConcyclic(A,S,D,E))}\end{center}
and  \begin{center}\texttt{ProveDetails(AreConcyclic(A,S,D,E))},\end{center}
yielding in both cases just the declaration that the statement is true. See items $a, b$ in the input column in the Figure \ref{fig2} or in Figure \ref{fig3}. Notice that Figure \ref{fig3} displays the circle through $A,S,D,E$, even if the angle $\angle BAD >  90^\circ$. Indeed, we have not made any formal implementation of such restriction, and this implies the statement holds even without such conditions. Thus, we can say we have already proved an extended version of the given problem.
 
 Let us conclude by  remarking two more things: one, that the internal proof is mathematically rigorous, dealing with symbolic computation algorithms (not through a  numerical or probabilistic approach); two, that the complexity of the involved computation has made impossible to output the list of associated degeneracy conditions (i.e.~geometric situations that must be avoided for the statement to hold true, like having $ABCD$ aligned) that could have been displayed (in simpler cases) on the output of the \texttt{ProveDetails} command.
 
 \section{Generalizing the Problem}
 
\begin{figure}
 \begin{center}
\includegraphics[width=0.8\linewidth]{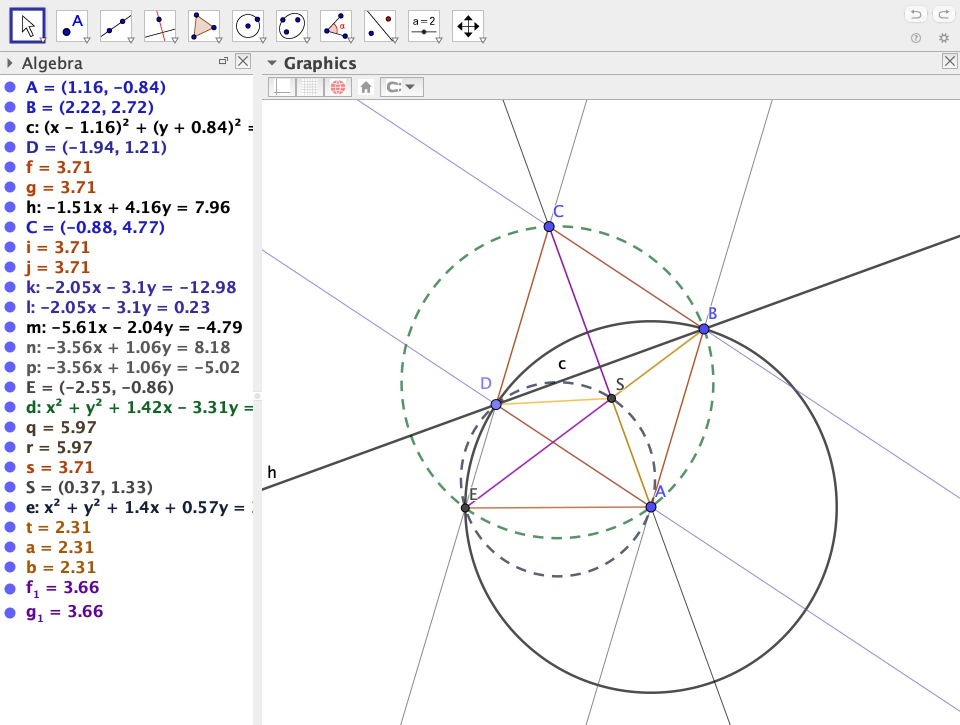}
\caption{Confirming the truth of the statement in Problem 2, and many others, through the \textit{Stepwise-discover} command.} 
\label{fig4}
\end{center}
\end{figure}

Conceptually speaking, GeoGebra Discovery is prone to offer tools that help, not only to check the truth/falsity of a given assertion, but to automatically test the verification of a large collection of properties among the elements that the user is adding in the construction of the geometric figure.  That is, to help the user \textit{discovering} different properties holding under the given hypotheses. Figure \ref{fig4} shows, in colors,  different properties that the program has discovered, after enabling the \texttt{StepwiseDiscovery} command, along each of the steps towards the construction of the rhombus, of the points $E, S$, etc.

For instance, just after introducing free points $A,B$ and point $D$ in the circle $c$, GeoGebra Discovery tells the user that segments $f=DA$, $t=BA$ have equal length, a trivial result, since $A$ is the center of circle $c$ and $B,D$ are points on $c$.  Perhaps more interesting is to learn that circle $d$ contains points $A,B,C,E$, i.e.~that these points are concyclic; or that  $t=BS$, $a=AS$, $b=DS$ have equal length, a couple of  non-trivial results that GeoGebra Discovery outputs automatically towards the end of the construction, without being asked to consider such specific relations.

\begin{figure}
 \begin{center}
\includegraphics[width=\linewidth]{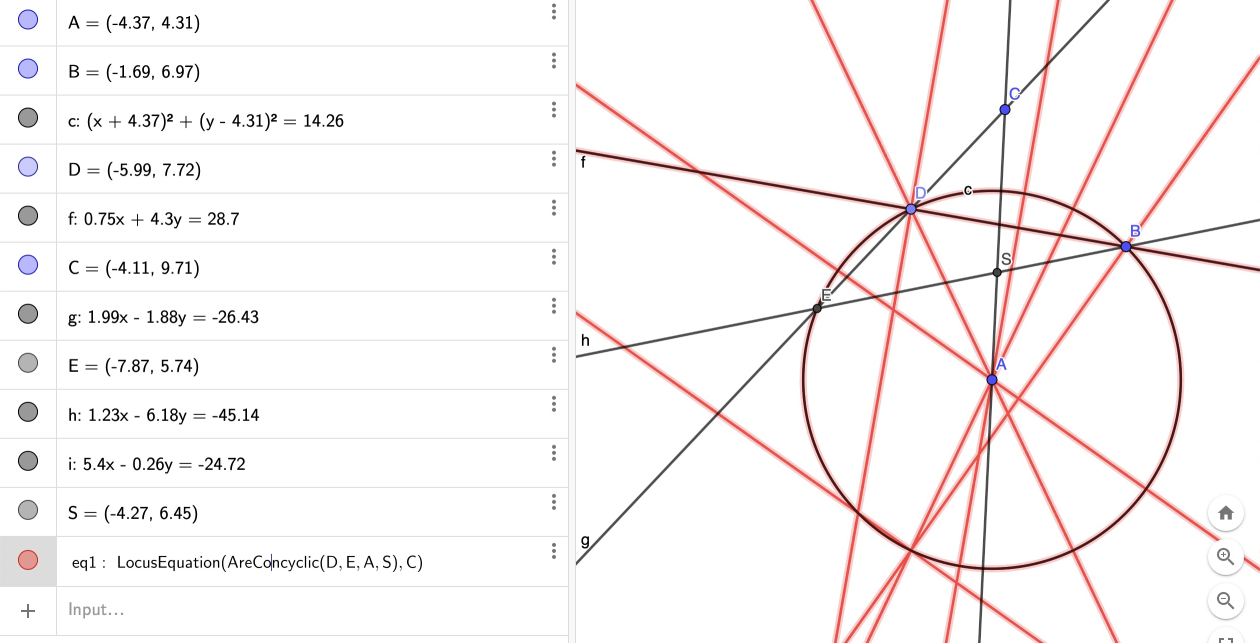}
\caption{Locus of $C$ (in red color) for the concyclicity of $A,S,D,E$, assuming only that $B,D$ are in a circle centered at $A$.} 
\label{fig5}
\end{center}
\end{figure}

We would like to focus on some other way, perhaps less automatic but more relevant for the user,  of \textit{discovering} results with GeoGebra Discovery. Namely, let us assume the user is interested in a precise question: finding necessary conditions for the converse of the given Problem 2. This can be addressed through the \texttt{LocusEquation(AreConcyclic(D,E,A,S),C)} command. Figure \ref{fig5} shows the output of this command, a very complicated degree-10-equation that (seems) to be the product of eight lines and the circle $c$.  The locus described by this equation includes all positions of a free point $C$ such that  $A,S,D,E$ are concyclic (where $A,S,D,E$ are defined repeating the construction of the previous figures, except for point $C$, that here is just a free point, not the mirror of $A$ with respect to the line $BD$).

Let us remark that the output is a numerical equation, with coefficients depending on the coordinates of the free points $A, B, D$ in the construction. So, it is neither a symbolic geometric object, nor an object we could enter in the ART to check if placing $C$ over some of the components implies $A,S,D,E$ are concyclic. We would have to learn how to build the locus starting from $A, B, D$, we would have to find some intrinsic geometric description of the locus, say, the analysis and discussion of the different lines.

\begin{figure}
 \begin{center}
\includegraphics[width=\linewidth]{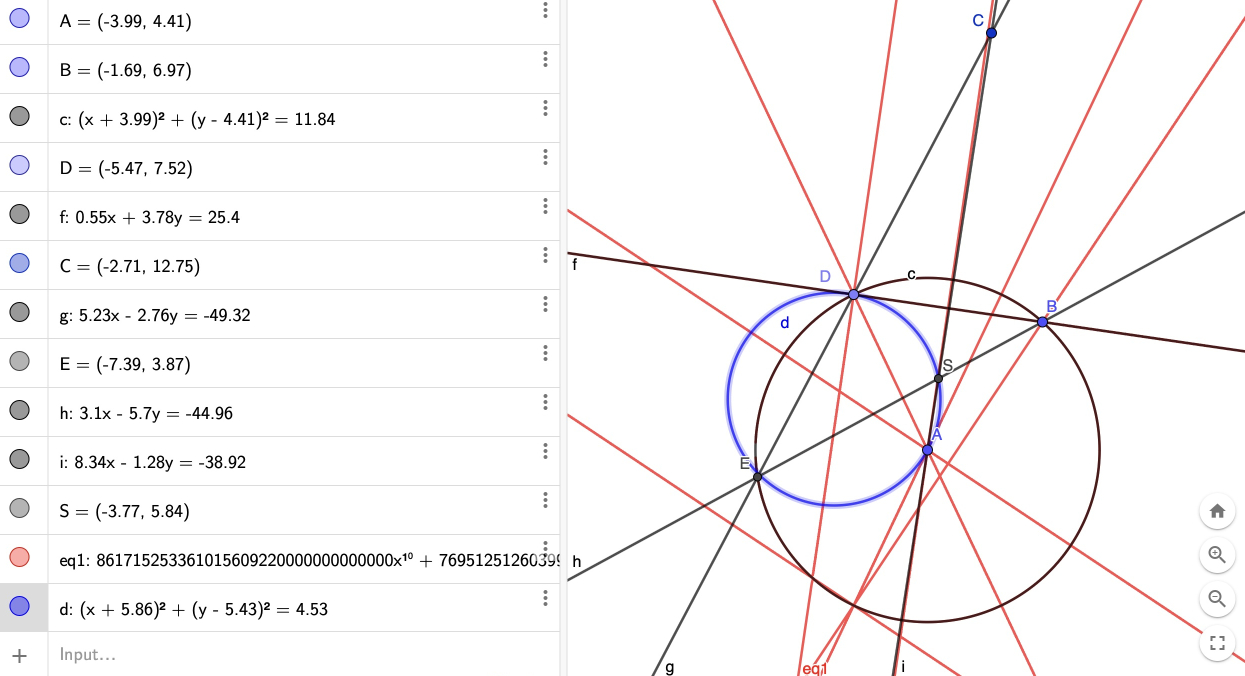}
\caption{Visually checking the validity of the locus of $C$ for the bisector line of $BD$.}
\label{fig6}
\end{center}
\end{figure}

\begin{figure}
 \begin{center}
\includegraphics[width=0.8\linewidth]{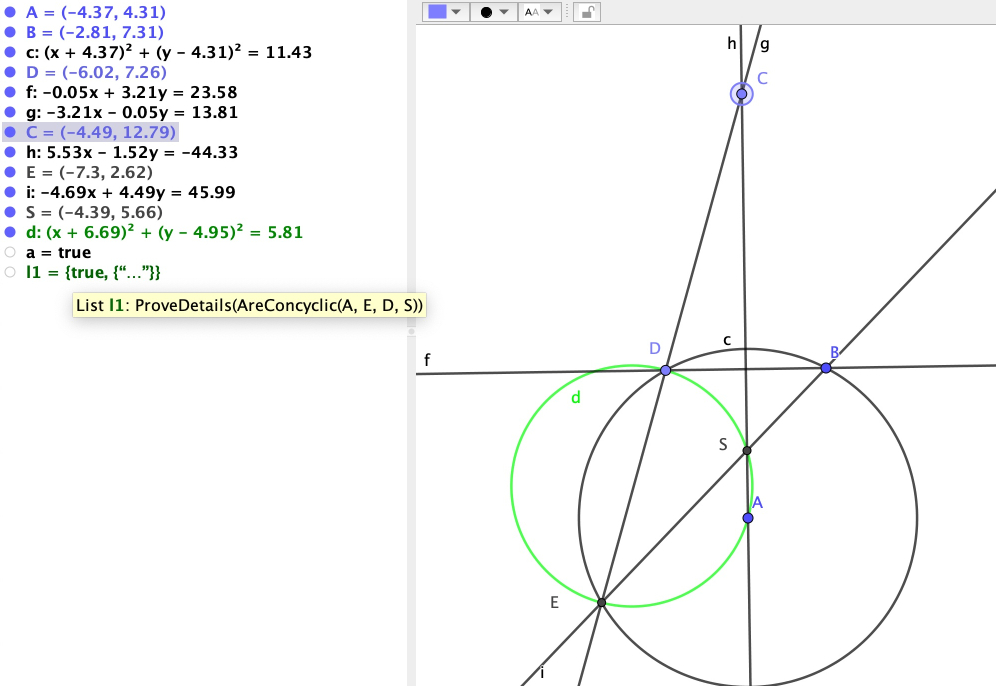}
\caption{Proving the validity of Problem 2 extended by placing $C$ just on  the bisector line of $BD$.} 
\label{fig7}
\end{center}
\end{figure}

In this particular case is not difficult to state that placing $C$ in the circle yields a degenerate case (since then $C=E=S$). And the same happens for several of the lines, except for the line that is perpendicular to $BD$, see Figure \ref{fig6} and, for a rigorous verification, Figure \ref{fig7}, displaying the original construction, but now $ABCD$ forms a \textit{kite}, since $C$ is placed anywhere on the bisector line of $BD$.

 Leaving aside the curious fact that, using GeoGebra Discovery ART we have been able to solve, and to generalize, an Olympiad Problem, we consider it is important  to reflect on some other consequences---on the educational context---of the sequence of facts we have described in this section.
 
 Indeed, it is well known the use of locus computation as a relevant methodology in mathematics education. For example, let us refer to the recent chapter  \cite{MPT}, where the authors describe an experience involving over 200 secondary education students from Sicily (Italy), analyzing in detail the impact of GeoGebra on the performance of the students that  had to find and to describe different geometric loci.
 
 We have realized (see \cite{BRT})  that the use of GeoGebra Discovery ART for accomplishing the same tasks, would imply a substantial methodological change: looking for most of the considered loci would turn out to be quite trivial, if it is just required finding the equation or displaying the graph of the locus. On the other hand, we think that the considered rich learning environment remains if the proposed activities  focus, instead,  on the description of the intrinsic geometric features of the obtained locus, as well as on their construction, two tasks that GeoGebra Discovery is not able to address automatically, but can contribute to its achievement.
 
In this context, we consider that the extended version of Problem 2 we have considered in this section, provides another excellent environment for exploring geometric locus with the help of automated reasoning tools, showing both their limitations as well as their useful features.
 
 \section{Grading Problem 2}

Very recently we have proposed an algorithmic way to associate, to each
geometry statement, a grade that intends to estimate its difficulty or
interest. It is yet a proposal in a very initial state---although already
implemented in the last version of GeoGebra Discovery, through the
\texttt{ShowProof} command---that, roughly speaking, computes (a bound
on) the degree of the polynomials $g_i$ expressing  the thesis polynomial
$T$ as a combination of the hypotheses polynomials $h_i$,
i.e.~$T=\sum_{i}g_i \cdot h_i$.

\begin{figure}
 \begin{center}
\includegraphics[width=\linewidth]{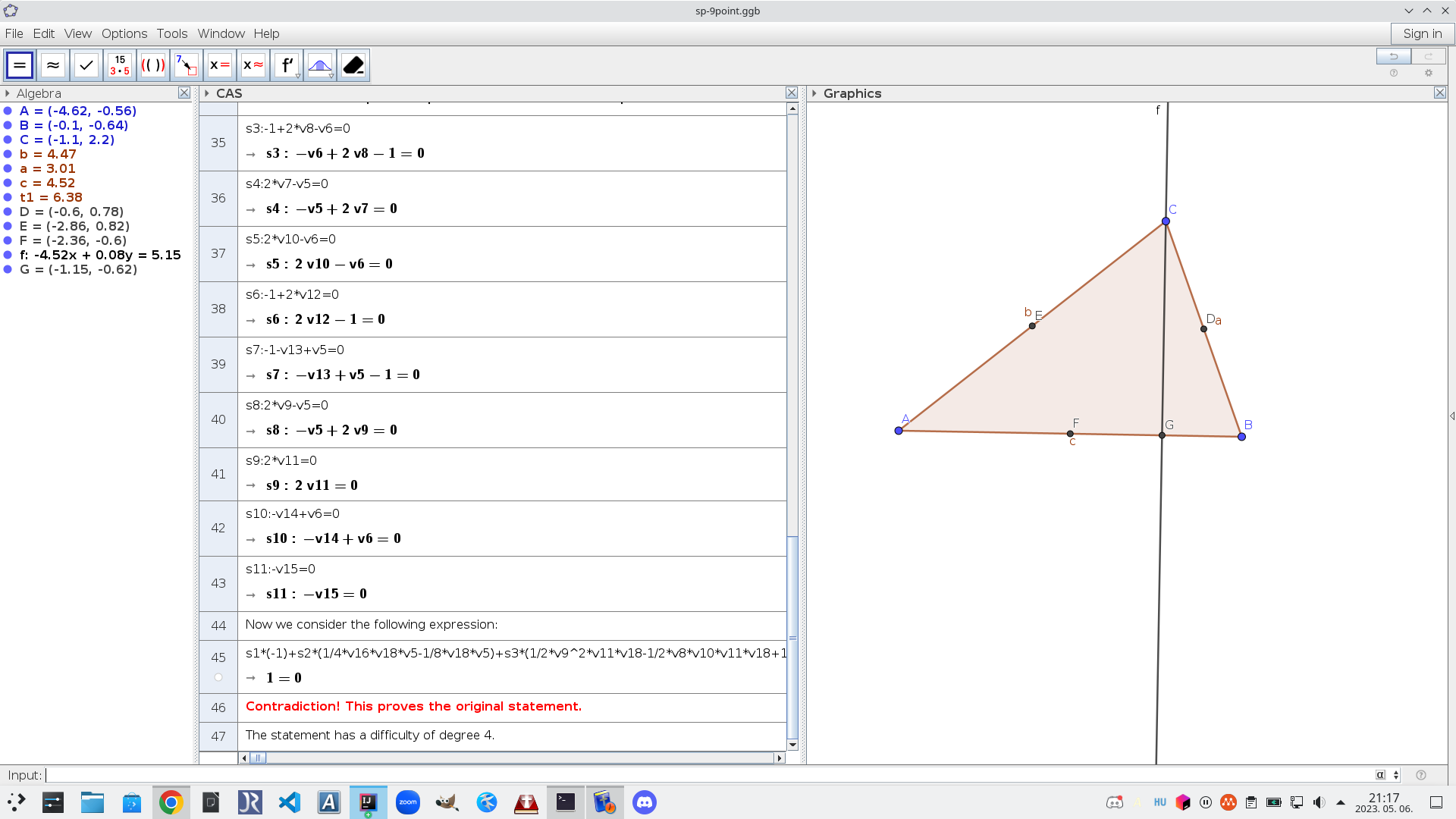}
\caption{Estimating the complexity of  proving the concyclicity of the midpoints of the sides and the feet of a height in a triangle.} 
\label{fig8}
\end{center}
\end{figure}

\begin{figure}
 \begin{center}
\includegraphics[width=\linewidth]{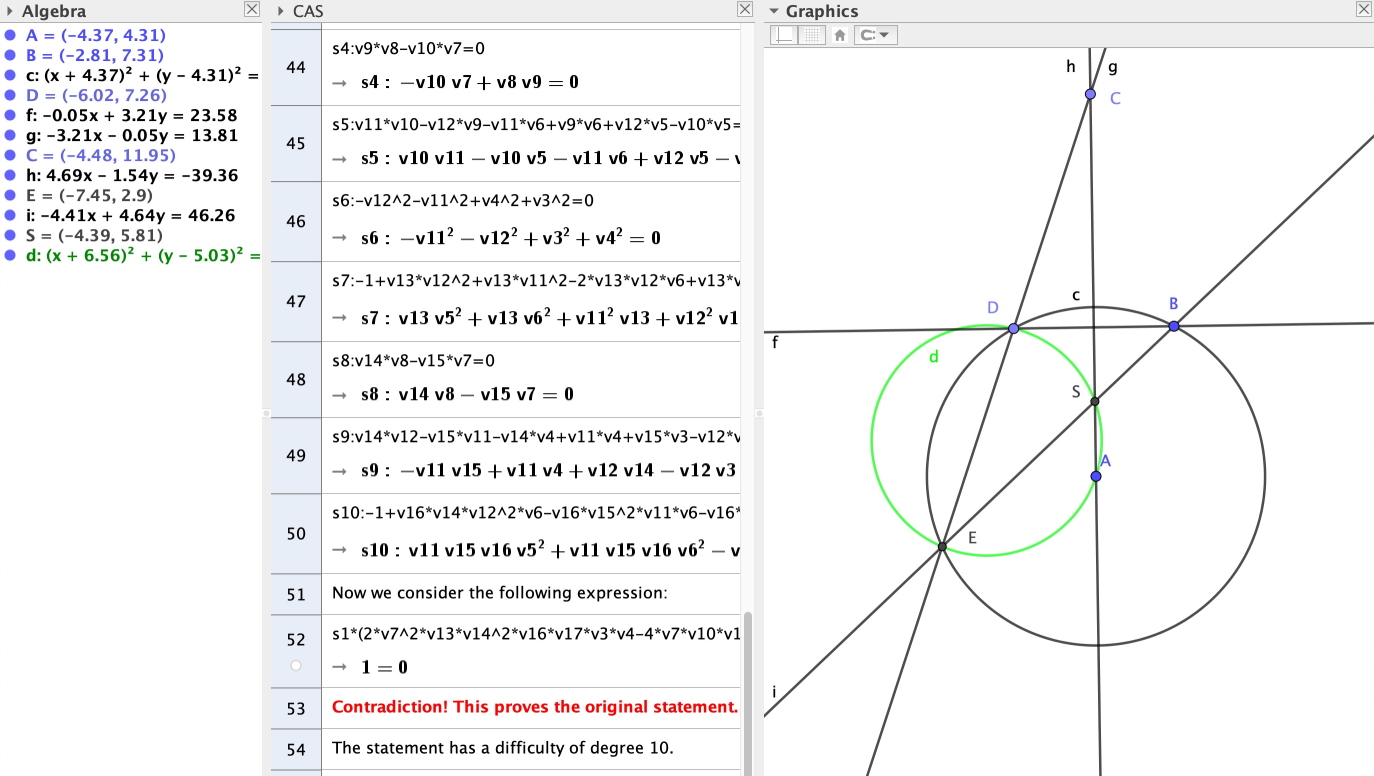}
\caption{Estimating the complexity of  Problem 2.} 
\label{fig9}
\end{center}
\end{figure}

We refer to \cite{KRV23}  for further details and examples.  Let us just mention here that most classical, elementary theorems (e.g.~Pythagoras, Intersection of Medians, Intersection of Heights, etc.) get grades 1 or 2. Or a partial formulation of the 9-point circle theorem gets complexity 4 (see Figure \ref{fig8}) while this Problem 2 of the Austrian Mathematical Olympiad has got grade 10!  See Figure \ref{fig9}.

To obtain this grade, GeoGebra Discovery internally  computes  a Gr\"obner basis of the hypotheses ideal (with respect to a certain degree ordering: the impact of the choice of order is still under study), with 13 elements. This computation outputs, as well, the expression of each element of the basis in terms of the hypotheses, with coefficients of degree bounded by 8. Then the expression of the thesis as a combination of the elements in the Gr\"obner basis is computed. Let us remark that in this formula, the degree of the multiplier polynomials is bounded by 3, but considering more precisely the sum of the involved degrees (of the polynomial multiplying  a certain element of the Gr\"obner basis times the maximum of the degrees of the polynomials expressing this element in terms of the hypotheses), the bound is limited to 10, a number that we think is adequate to be associated to a Mathematical Olympiad Problem of an intermediate level of difficulty.

 \section{Conclusions}
  In our communication we have illustrated, considering just a single problem from the Austrian Mathematical Olympiad 2023,  several facts concerning the use of GeoGebra Discovery in the classroom:  
  
  \begin{enumerate}
  \item  the ability of GeoGebra Discovery Automated Reasoning Tools (ART) to immediately solve a problem presented at a regional Mathematics Olympiad, that the recent GeoGebra ART complexity measure ranks quite highly, 
  \item the use of GeoGebra Discovery as a decisive auxiliary tool to develop and confirm new, non-trivial, conjectures, such as the generalization of the proposed problem,
  \item the need to change the methodological focus  when working with locus computation in the classroom with  Dynamic Geometry programs, from finding equations and displaying its graph, to analyzing and obtaining  the geometric  characteristics of the involved locus, and its construction, by using GeoGebra Discovery ART.
  \end{enumerate}
  
  The opportunity to consider simultaneously all these items around a single problem, is probably the most relevant contribution of this communication.


\begin{thebibliography}{8}

\bibitem{KRV22} Kov\'acs, Z.; Recio, T.; V\'elez, M. P.: Automated reasoning tools with GeoGebra: What are they? What are they good for? In: P. R. Richard, M. P. V\'elez, S. van Vaerenbergh (eds): Mathematics Education in the Age of Artificial Intelligence: How Artificial Intelligence can serve mathematical human learning. Series: Mathematics Education in the Digital Era, Vol. 17, pp. 23--44. Springer Cham (2022). \doi{10.1007/978-3-030-86909-0\_2}

\bibitem{ART} Ari\~no-Morera, B.; Recio, T.; Tolmos, P.: Olympic geometry problems: human vs.~machine. Communication to the CADGME (Digital Tools in Mathematics Education) 2022 Conference.  Abstracts available at \url{https://drive.google.com/file/d/1qF4ceMg6gNklOPa1JVkgKND1dOqNmyka/viewpp}

\bibitem{AM} Ari\~no-Morera, M. B.:  GeoGebra Discovery at EGMO 2022. Revista Do Instituto GeoGebra Internacional De S\~ao Paulo, 11(2), 005-016.2022. \doi{10.23925/2237-9657.2022.v11i2p005-016}

\bibitem{KRV23} Kov\'acs, Z.; Recio, T.; V\'elez, M. P.: Showing proofs,
assessing difficulty with GeoGebra Discovery. Communication to  the ADG
(Automated Deduction in Geometry) Conference, Belgrade, 2023.
\url{https://adg2023.matf.bg.ac.rs/downloads/slides/ShowingProofsZoltanRecioVelez.pdf}
\doi{10.13140/RG.2.2.31885.31205}

\bibitem{MPT} Mammana, M.F.; Pennisi, M.; Taranto, E.:  Teaching Intriguing Geometric Loci with DGS. In: Aldon, G., Hitt, F., Bazzini, L., Gellert, U. (Eds). Mathematics and Technology. Advances in Mathematics Education. Springer, Cham (2017). \doi{10.1007/978-3-319-51380-5\_26}

\bibitem{BRT} Ari\~no-Morera, B.; Recio, T.; Tolmos, P.: Teaching Intriguing Geometric Loci with GeoGebra Discovery. Communication to the CADGME (Digital Tools in Mathematics Education) 2023 Conference. \url{https://sites.google.com/view/cadgme2023/program}


\end{thebibliography}
\end{document}